\documentclass[twocolumn]{aastex631}

\usepackage{CJK}

\begin{document}
\begin{CJK*}{UTF8}{bsmi}

\title{Detection of Methane in the Closest Extreme Metal-poor T Dwarf \\ WISEA\,J181006.18$-$101000.5\footnote{Based on observations made with the Gran Telescopio Canarias.}}

\author[0000-0001-5392-2701]{Jerry J.-Y.\ Zhang (章俊龑)}
\affiliation{Instituto de Astrof\'isica de Canarias (IAC), 
Calle V\'ia L\'actea s/n, E-38200 La Laguna, Tenerife, Spain}\affiliation{Departamento de Astrof\'isica, Universidad de La Laguna (ULL), E-38206 La Laguna, Tenerife, Spain}

\author[0000-0002-3612-8968]{Nicolas\ Lodieu}
\affiliation{Instituto de Astrof\'isica de Canarias (IAC), 
Calle V\'ia L\'actea s/n, E-38200 La Laguna, Tenerife, Spain}\affiliation{Departamento de Astrof\'isica, Universidad de La Laguna (ULL), E-38206 La Laguna, Tenerife, Spain}

\author[0000-0002-1208-4833]{Eduardo L. Mart\'in}
\affiliation{Instituto de Astrof\'isica de Canarias (IAC), Calle V\'ia L\'actea s/n, E-38200 La Laguna, Tenerife, Spain}\affiliation{Departamento de Astrof\'isica, Universidad de La Laguna (ULL), E-38206 La Laguna, Tenerife, Spain}

\author[0000-0001-6172-3403]{Pascal Tremblin}
\affiliation{Université Paris-Saclay, UVSQ, CNRS, CEA, Maison de la Simulation, 91191 Gif-sur-Yvette, France}

\author[0000-0001-5664-2852]{Mar\'ia Rosa Zapatero Osorio}
\affiliation{Centro de Astrobiolog\'ia (CAB), CSIC-INTA, Camino Bajo del Castillo s/n, 28692 Villanueva de la Ca\~nada, Madrid, Spain}

\author[0000-0002-5086-4232]{V\'ictor J. S. B\'ejar}
\affiliation{Instituto de Astrof\'isica de Canarias (IAC), Calle V\'ia L\'actea s/n, E-38200 La Laguna, Tenerife, Spain}\affiliation{Departamento de Astrof\'isica, Universidad de La Laguna (ULL), E-38206 La Laguna, Tenerife, Spain}

\author[0000-0003-2150-0787]{Nikola Vitas}
\affiliation{Instituto de Astrof\'isica de Canarias (IAC), Calle V\'ia L\'actea s/n, E-38200 La Laguna, Tenerife, Spain}\affiliation{Departamento de Astrof\'isica, Universidad de La Laguna (ULL), E-38206 La Laguna, Tenerife, Spain}

\author[0000-0001-5452-2056]{Bartosz Gauza}
\affiliation{Janusz Gil Institute of Astronomy, University of Zielona G\'ora, Lubuska 2, 65-265 Zielona G\'ora, Poland} 

\author[0000-0002-7615-4028]{Yakiv\ V.\ Pavlenko}
\affiliation{Main Astronomical Observatory, Academy of Sciences of the Ukraine, 27 Zabolotnoho, Kyiv 03143, Ukraine}\affiliation{Instituto de Astrof\'isica de Canarias (IAC), Calle V\'ia L\'actea s/n, E-38200 La Laguna, Tenerife, Spain}

\author[0000-0003-3767-7085]{Rafael Rebolo}
\affiliation{Instituto de Astrof\'isica de Canarias (IAC), Calle V\'ia L\'actea s/n, E-38200 La Laguna, Tenerife, Spain}\affiliation{Departamento de Astrof\'isica, Universidad de La Laguna (ULL), E-38206 La Laguna, Tenerife, Spain}

\correspondingauthor{Jerry Zhang}
\email{jzhang@iac.es}



\begin{abstract}

 WISEA\,J181006.18$-$101000.5 (WISE1810) is the nearest metal-poor ultracool dwarf to the Sun. It has a low effective temperature and has been classified as extreme early-T subdwarf. However, methane, the characteristic molecule of the spectral class T, was not seen in the previous low-resolution spectrum. 
Using the 10.4-m Gran Telescopio Canarias, we collected a high-quality $JHK$-band intermediate-resolution $R\approx5000$ spectrum of WISE1810, in which a $17\pm6$ ppm of methane is clearly detected, while carbon monoxide is absent. Based on customly computed ATMO2020++ model, we estimated an effective temperature of $1000\pm100$\,K, a high surface gravity of $\log g = 5.5 \pm 0.5$\,dex, a carbon abundance [C/H]\,$=-1.5\pm0.2$ dex, inferring [Fe/H]\,$=-1.7\pm0.2$ dex. Potassium is not seen in our data, 
and the upper limits of pseudo-equivalent width of $J$-band atomic lines are at least 25 to 60 times weaker than those measured from solar-metallicity early-T counterparts. We measured a heliocentric radial velocity of $-83\pm13$ km\,s$^{-1}$, inferring that WISE1810 is more likely a thick disk member.

\end{abstract}

\keywords{T dwarfs (1679), T subdwarfs (1680), Brown dwarfs (185), Metallicity (1031), Infrared spectroscopy (2285), Population II stars (1284)}


%
\section{Introduction}
\label{esdT_WISE1810:intro}

\begin{figure*}[htbp]
   \centering
   \includegraphics[width=0.85\linewidth, angle=0]{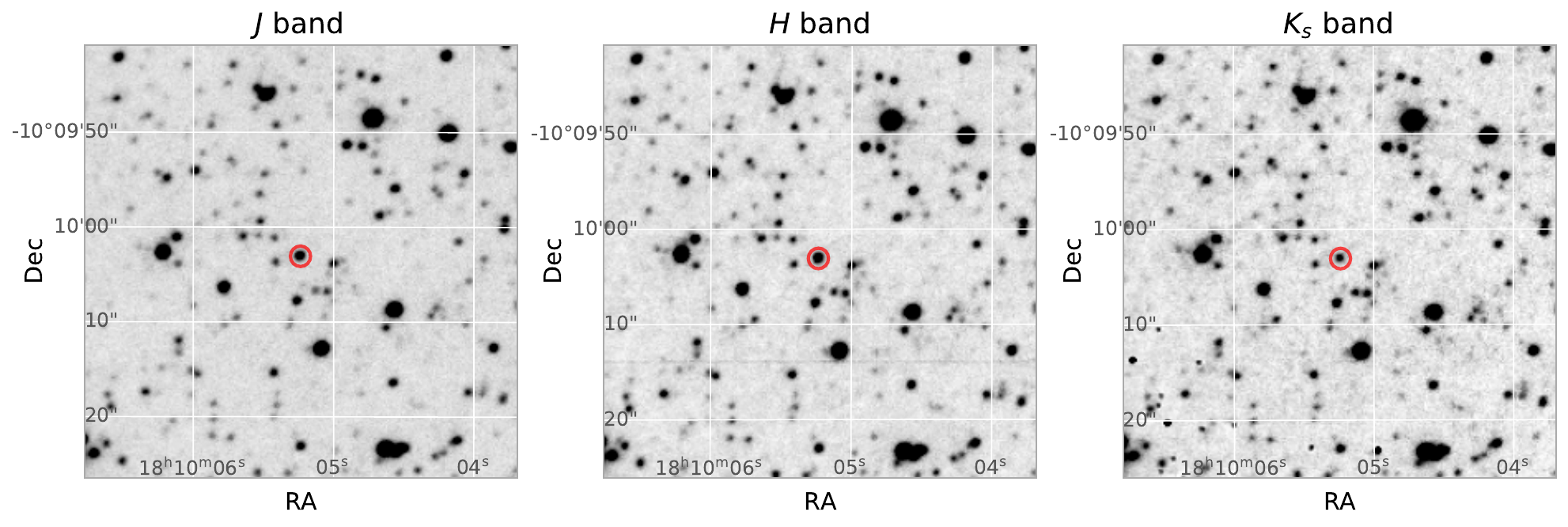}
   \caption{
   Images of the position of WISE1810 on 21 July 2024 from GTC/EMIR in the $J$, $H$ and $K_s$ bands. The object is marked with a red circle. The field of view of each image is 45\arcsec$\times$45\arcsec. North is up and East left.
   }
   \label{fig_WISE1810:WISE1810images}
\end{figure*}

The physical properties of substellar objects evolve with time. The in-depth characterization of molecules in brown dwarf atmospheres over a wide range of ages and metallicities is critical to trace their evolutionary path and constrain their physical properties. Since the first discoveries of the first two unambiguous brown dwarfs \citep{rebolo1995discovery,nakajima1995discovery}, the field has grown up significantly with thousands of substellar objects with diverse properties: mass, age, and atmospheric composition \citep{Kirkpatrick2024IMF3600}. However, very little is known at metallicity an order of magnitude below solar for temperatures cooler than 1000\,K\@.

There are a few slightly metal-poor T-type dwarfs belonging to the thick disk or halo announced to date. Three isolated metal-poor subdwarfs have been discovered \citep{burgasser2003esdL7_2M0532,murray_burningham2011blueT,burningham2014T6.5} along with three wide companions to stars with well-determined metallicities \citep{burningham2010cool_binary_sdL,scholz2010ULAS1416,pinfield2012sdT8,mace2013Wolf1130C}. In the past years, a dozen new T subdwarfs have been spotted in the Wide-field Infrared Survey Explorer \citep[WISE;][]{wright2010WISE} with peculiar colours \citep{greco2019neowise,meisner2020extremecoldBD,meisner2021esdT,brooks2022esdT,burgasser2024esdT_class}. Nonetheless, the first two cases of extreme T subdwarfs (esdT) with metallicities likely below $-$1.0\,dex have been announced by \citet{Schneider2020W0414_W1810}, WISEA\,J181006.18$-$101000.5 (WISE1810) and WISEA\,J041451.67$-$585456.7, with infrared photometry and spectroscopy not well reproduced by any model or observed T dwarf templates.

In this paper, we focus on WISE1810, the nearest T subdwarf to the Sun, located at $8.9\pm0.7$\,pc \citep{lodieu2022W1810}. 
There is a discrepancy in the effective temperature for WISE1810: from $1300\pm100$\,K \citep{Schneider2020W0414_W1810} to $800\pm100$\,K \citep{lodieu2022W1810}; as well as its spectral type determinations: from esdT0 \citep{Schneider2020W0414_W1810} to esdT3 \citep{burgasser2024esdT_class}. Based on previous low-resolution ($R\approx2600$) near-infrared (NIR) spectra and mid-IR photometry, methane (CH$_4$) was not detected, suggesting a significantly chemically modified atmosphere \citep{lodieu2022W1810, burgasser2024esdT_class} and also rising question to the spectral classification, as CH$_4$ is a defining characteristic molecule of the T type. Hence, a spectroscopy with higher resolution and better quality is highly desirable to assess the presence of methane in the atmosphere of this unique object.


\section{Observations}
\label{esdT_WISE1810:Obs}

\subsection{Photometry}
\label{esdT_WISE1810:Obs_phot}
We collected new NIR photometry with the Espectrografo Multiobjeto Infra-Rojo \citep[EMIR;][]{garzon2022EMIR} mounted on the Nasmyth-A focus of the Gran Telescopio Canarias (GTC) as part of programme GTC73-24A (PI Lodieu). Since December 2023, EMIR has been equipped with a new 2048$\times$2048 pixel Teledyne HAWAII-2RG HgCdTe detector offering a 6.67\arcmin$\times$6.67\arcmin\,field of view with a pixel scale of 0\farcs2. 

We observed WISE1810 using The Two Micron All Sky Survey \citep[2MASS; ][]{skrutskie2006_2MASS} $JHK_{s}$ filters on the night of 21 July 2024 in visitor mode under clear skies, 0\farcs6 seeing, and full moon. For each filter, we collected images with one 7-point dithering pattern with an offset of 10\arcsec\, and 10s individual exposure (Table~\ref{tab_WISE1810:tab_logs}). 

We reduced the images with the official EMIR pipeline PyEMIR \citep{Cardiel2019pyemir}. The pipeline did bias and flat field corrections for each individual image and constructed the sky of each dithering point by combining all the images at the other six positions. Then the sky was subtracted and all the images were stacked applying the offset between each dithering point.

%
%
\begin{table}[htbp]
\centering
 \caption{New WISE1810 astometry and NIR photometry.}
\footnotesize
 \begin{tabular}{ll}
 \hline
MJD &  60512.91 \\
$\alpha,\delta$ (\degr)&$272.5218311\pm0.000013$,  $-10.1675257\pm0.000013$\\
$\alpha,\delta$ &$18^{\mathrm{h}}10^{\mathrm{m}}5\fs239\pm 0\fs003$, $-10\degr10\arcmin3\farcs093\pm0\farcs046$\\
$\mu_\alpha\cos\delta,\mu_\delta$&$-1038.1\pm4.0$ mas\,yr$^{-1}$, $-243.2\pm4.0$ mas\,yr$^{-1}$\\
$J_{\mathrm{2MASS}}$   & $17.281\pm0.032$\,mag (Vega) \\
$H_{\mathrm{2MASS}}$  & $16.446\pm0.031$\,mag  (Vega) \\
${K_s}_{\mathrm{2MASS}}$  & $16.911\pm0.057$\,mag  (Vega) \cr
%
\hline
 \label{phottable_W1810}
 \end{tabular}
\end{table}

\vspace{-6mm}

WISE1810 was identified based on its proper motion and previous positions \citep{lodieu2022W1810}. Figure~\ref{fig_WISE1810:WISE1810images} shows the position of WISE1810, which lies on a relatively empty region in all the three bands. 
We astrometrically solve the images using astrometry.net and measured the $J$-band position of WISE1810 using the \textit{imcentroid} task of the Image Reduction and Analysis Facility \citep[IRAF;][]{tody1986iraf,tody1993iraf}. We took advantage of the longest available baseline of 14.07 years between the current epoch and that of the Galactic Plane Survey J-band image \citep{lucas2008ukidss_gps} and updated its proper motion (Table~\ref{phottable_W1810}).

We performed an aperture photometry using the {\tt{Photutils}} package \citep{photutil} using apertures with a radius of 0\farcs9 $\approx$ 1.5 FWHM, sky annuli with an inner and outer radius of 3\farcs5 and 5\farcs5, respectively. The calibration was done using 2MASS point sources \citep{cutri2003_2MASSpoint} in the field,
 and the errors are calculated as the quadratic sum of the source's Poisson noise, the sky fluctuation, and the dispersion of the photometric calibration (Table~\ref{phottable_W1810}). The results are consistent with the photometry published by \citet{Schneider2020W0414_W1810} and \citet{lodieu2022W1810}, inferring that WISE1810 has long-term (three years to one and a half decades) variability $\lesssim 0.1$\,mag in the $JH$ bands, and $\lesssim 0.2$\,mag in the $K$ band.

%
%
%
\subsection{Spectroscopy}
\label{esdT_WISE1810:Obs_spec}

We also obtained an improved NIR spectrum of WISE1810 from 21 to 25 July 2024 for five consecutive nights in visitor mode. We used the $JHK$ grisms with a slit of 0\farcs6 to yield spectral resolutions of 5000, 4500, and 4000 in the 11700--13300, 15200--17700, and 20300--23700\,\AA{} intervals, respectively.
We dedicated the first and fifth night to the $J$ grism, the second night to $H$, and the third and fourth nights to $K$ (Table \ref{tab_WISE1810:tab_logs}). The on-source integration times were 3.6h, 3.6h, 7.2h for the $JHK$ grisms, respectively. We observed a telluric standard star, HIP\,88374 \citep[B9IV; ][]{shokry2018Bestars} at the beginning and end of the first four nights just before and after WISE1810, i.e.\ close in time and airmass. 

We set the individual exposure time for WISE1810 to 360s and created an ABBA pattern shifting the object along the slit by 10\arcsec\, to remove the sky contribution. The closest source to WISE1810 lies at about 5\arcsec\,to the south along the slit. The numbers of repeats depend on the night. The individual exposure time for the telluric standard were set to 10s, then with a ABBA pattern.

We also observed 2MASS\,J20304235$+$0749358, a bright T1.5 dwarf \citep[2MASS2030, $J$\,=\,14.22 mag;][]{best2013PS1_WISE_LT} as a RV standard with an RV of 21.2 km\,s$^{-1}$ \citep{hsu2021BDKP} in the $J$-band with the same configuration. 
We set the individual exposures to 200s with one ABBA loop, yielding a total 
exposure time of 800s. 


We reduced the EMIR spectra with PyEMIR to produce a final stacked 2D spectrum. The pipeline rectified the spectra, calibrated the wavelength in vacuum using OH airglow lines, subtracted the sky background using the adjacent A/B image and stacked all the sky-subtracted images knowing the offset along the slit. The nominal wavelength steps are 0.77, 1.22, and 1.73\,\AA{}pix$^{-1}$ in the $JHK$ spectra, respectively.  We extracted the 1D spectrum using IRAF \textit{apall} task with manually selected aperture. We repeated the same process for the telluric standard, and masked all the prominent Paschen and Brackett lines in the spectrum. Finally, we divided the target spectrum by the masked telluric and then multiplied by the spectrophotometric standard spectra from ESO library\footnote{ \url{https://www.eso.org/sci/observing/tools/standards/IR_spectral_library.html}} \citep{pickles1998spectral_library}. The relative flux ratios among the three bands were calibrated using the first night photometric results\footnote{The GTC/EMIR filter information is available at SVO service \url{http://svo2.cab.inta-csic.es/theory/fps/index.php?mode=browse&gname=GTC&gname2=EMIR&asttype=}}.
We applied the same procedure to the RV standard.

\begin{figure*}
   \centering
    \includegraphics[width=\linewidth]{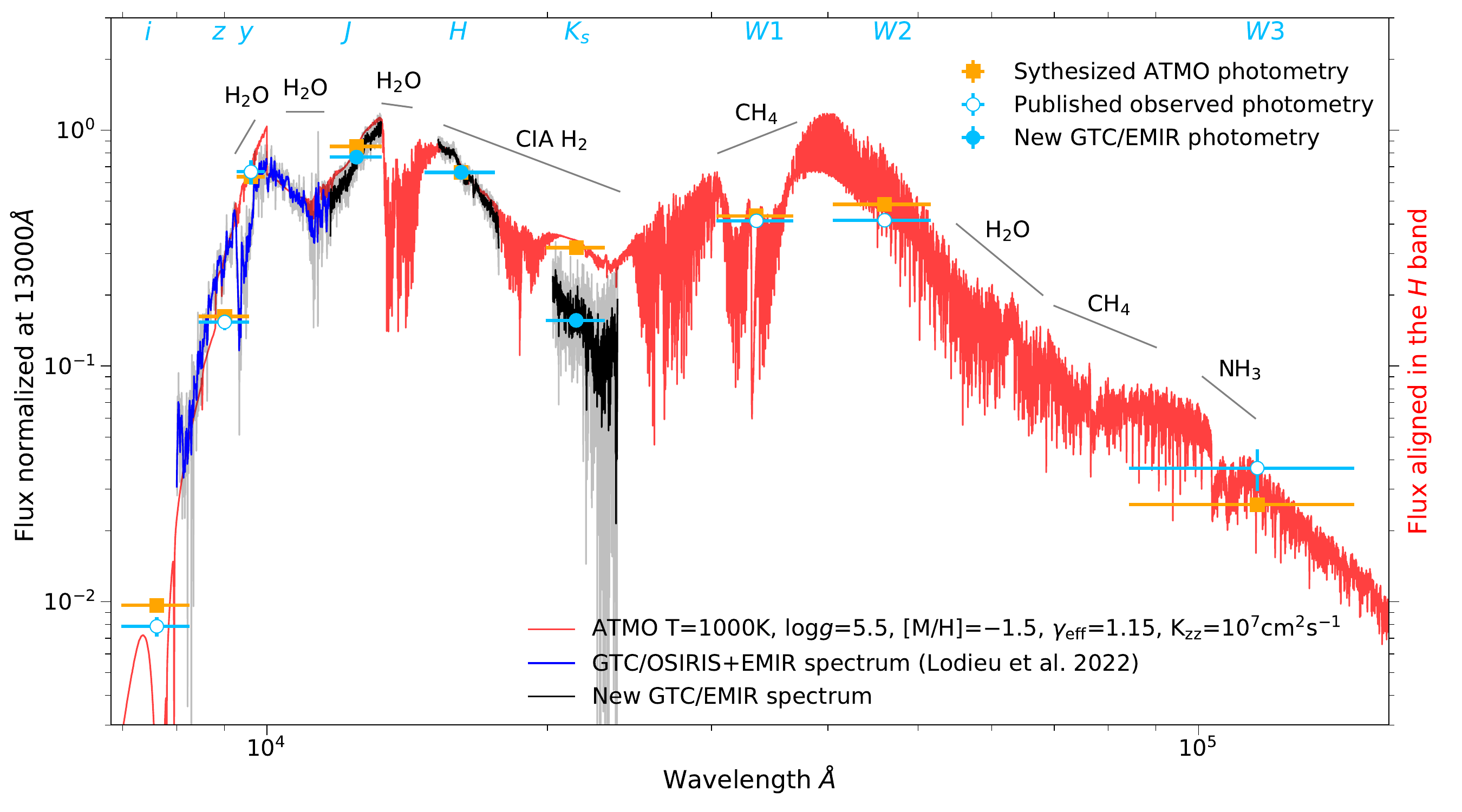}
   \includegraphics[width=\linewidth, angle=0]{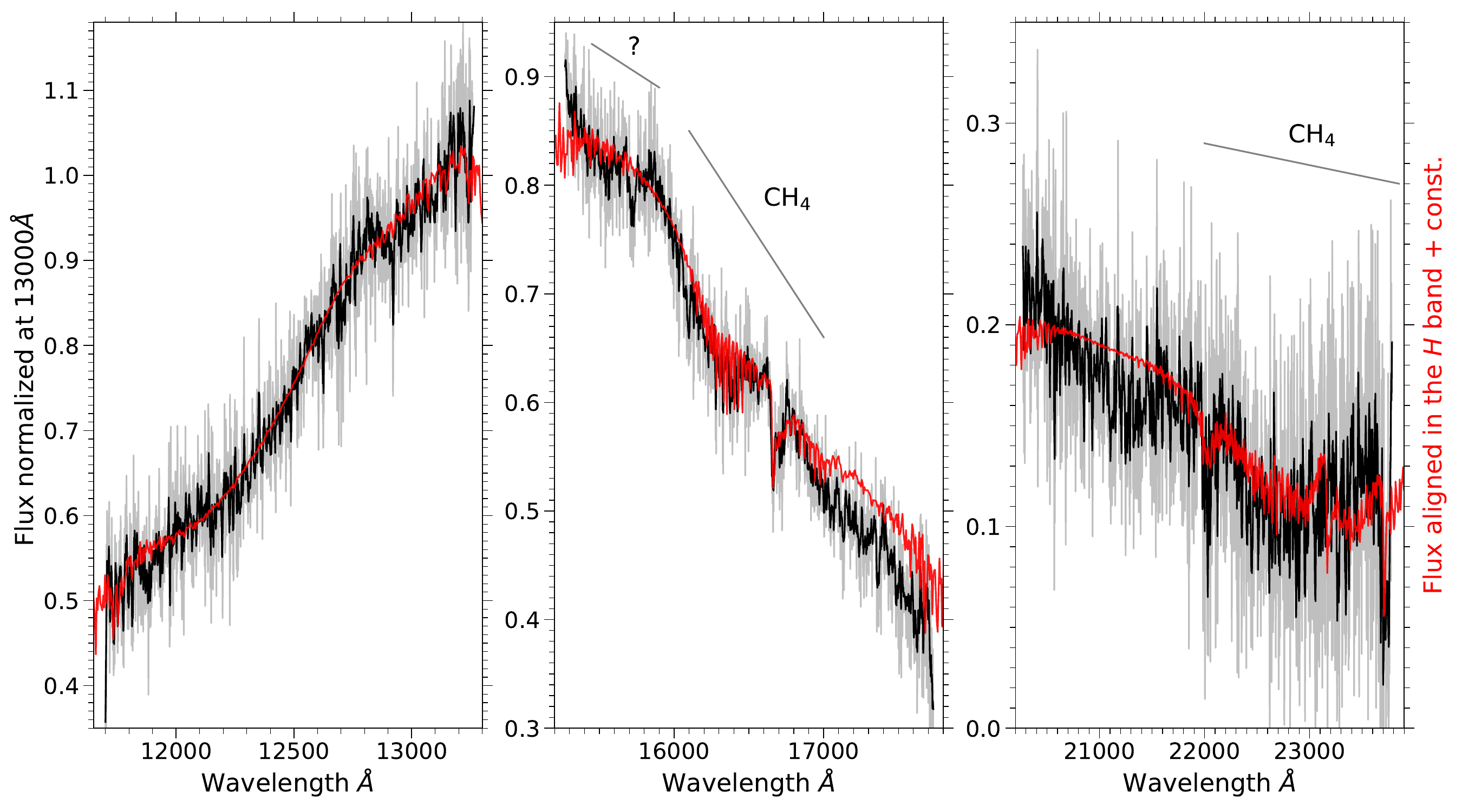}
   \caption{New WISE1810 $JHK$ spectrum normalized at 13000\,\AA{} compared with ATMO2020++ best-fit model matching in the $H$ band (red). Upper: the whole spectral energy distribution, combined with multi-band photometry, GTC/OSIRIS red optical spectrum (8000-10000 \AA{}) and  GTC/EMIR $YJ$ spectum (10000-11700 \AA{}) from \citet{lodieu2022W1810}; Lower: the $JHK$ bands, with model matched in the $H$ band but moved onto the observed spectra to compare features in the $J$ and $K$ bands. Note: the original data is plotted in light grey and solid black or blue lines are data smoothed by a factor of 7.
   }
   \label{fig_WISE1810:EMIRspec}
\end{figure*}

We estimated the signal-to-noise ratio (S/N) per pixel of the spectrum (Fig.~\ref{fig_WISE1810:EMIRspec}) using IRAF \textit{splot} task. The $JHK$ band has an average S/N of 15, 20, and 4, respectively.
Compared with the previous spectrum from \citet{Schneider2020W0414_W1810}, our spectrum has a higher resolution and finer quality (Fig.~\ref{compare}). In addition, our $K$-band spectrum differs by a factor of 2, which might be due to variability (not supported by the available photometry) but more likely to spectroscopic flux calibration issues.  

%
%


\section{Chemistry}
\label{esdT_WISE1810:chemistry}

\subsection{Model fit}
We combined our spectrum and optical spectrum, $YJ$-band spectrum, and photometry points from \citet{lodieu2022W1810} (Fig.~\ref{fig_WISE1810:EMIRspec}). We used the theoretical spectra computed from the ATMO2020++ grid of models \citep{leggett2021coldestSED,meisner2023coldoldBD} and degraded to the resolution of $R=3000$. The least squares method was used to find the best-fit model to constrain the effective temperature, the surface gravity, and the metallicity. Starting from a grid model, the temperature-pressure profile (left panel of Fig.~\ref{pressure-abundance}) is adjusted to improve the fit of the observed spectral energy distribution. Since the $YJ$-band flux was too high compared to the $HK$-band flux, the temperature gradient in the model atmosphere is reduced by reducing the effective adiabatic index $\gamma_{\mathrm{eff}}$ from 1.25 to 1.15 \citep{tremblin2015atmo}. 
The fitting provides an effective temperature of $T_{\mathrm{eff}}=1000$\,K, and a surface gravity of $\log(g)=5.5$ dex. The initial metallicity was at the lower limit of the grid ($-$1.0\,dex) produced too strong  molecular features in the synthetic spectrum. Therefore, in our adjusted model the metallicity was reduced down to [M/H]\,$=-$1.5\,dex. In the fitting procedure the relative abundances of all metals are scaled down for the same factor. The step size of the grid is 100\,K for the temperature, 0.5\,dex for the gravity, and 0.5\,dex for the metallicity.  As it is shown in the upper panel of Fig.~\ref{fig_WISE1810:EMIRspec}, the overall fit is satisfactory except for the excess flux in the $K$ band.  
These values agree with those of \citet{lodieu2022W1810} using the LOWZ model \citep{meisner2021esdT}: $800\pm100$\,K, $\log(g)=5.00\pm0.25$\,dex and metallicity $-1.5\pm0.5$\,dex. 

\subsection{Carbon}


The CH$_4$ absorption bands at 16000--16700\,\AA{} and at 22100\,\AA{} \citep{yurchenko2014CH4} are clearly detected, supporting the classification of WISE1810 as T-type instead of L-type \citep{zhangjerry2023optical_sdT}. 
On the other hand, the CO absorption band head starting at 22900\,\AA{} is not seen.
It strongly favors a lower effective temperature determination ($\lesssim$ 1200\,K) 
and suggests that the CO molecules are contained in hotter and deeper subatmospheric layers. 
According to the model, CO has abundance at least three orders of magnitude less than CH$_4$ in the pressure range $10^{-4}$ to $10^{4}$ bar, while the abundance of CO$_2$ is even lower (right panel of Fig.~\ref{pressure-abundance}). Unfortunately, this fact limits our ability to constrain the vertical mixing strength or determine whether the atmosphere is out of equilibrium.

The depth of the CH$_4$ features in the $HK$ bands is well reproduced by the model, as well as the $W1$-band flux which is dominated by the CH$_4$ absorption. Because the fit between the observations and the theoretical spectra is dominated by the carbon spectral features, the metallicity we adopted is primarily a measure of the relative abundance of carbon [C/H]\,=\,[M/H]\,$=-1.5$\,dex.  From the model and the observational uncertainties, we have a CH$_4$ abundance of $1.7 \times 10^{-5}\pm0.6 \times 10^{-5}$, or $17\pm6$ parts per million (ppm). The uncertainty of carbon abundance is constrained by the CH$_4$ and hence is $\pm0.2$\,dex, assuming that CH$_4$ is the sole carbon-bearing molecule and ignoring its isotopic variants which are very likely much less abundant. This value may serve as a metallicity proxy: for metal-poor thick-disk stars [C/Fe] $\simeq0.2$ dex \citep{nissen2014C_O_population}, hence WISE1810 would have a [Fe/H] of $-1.7\pm0.2$\,dex.



\subsection{Potassium}

Although K{\small{I}} lines are still too weak to be seen, we constrained their strength to be an order of magnitude lower than previous limits \citep{lodieu2022W1810}. One of them is blended by two lines at 11772.86 and 11776.06\,\AA{} at our spectral resolution; and the other doublet is at (12435.68, 12525.57)\,\AA{} \citep[][wavelengths in vacuum]{NIST_ASD}. We use the IRAF {\tt{splot}} task to estimate an upper limit of the pseudo-equivalent widths pEWs. For the first blended line, pEW\,$<$\,0.5\,\AA{}, which is at least 25 times less than those of solar-metallicity early-T dwarfs \citep{faherty2014cloud_t_g_luhman};  and is at least 10 times less than those of early-T subdwarf candidates \citep{burgasser2024esdT_class}. Both lines of the doublet have pEW\,$<$\,0.2\,\AA{}, which is at least 30 and 60 times weaker than those of solar-metallicity counterparts, respectively (Fig.~\ref{fig:Klines}). 
For most of T subdwarf candidates we do not have this doublet measured yet \citep{burgasser2024esdT_class}.

The low metallicity of WISE1810 probably accounts majorly for the non-detection of atomic potassium but a lower temperature could also boost this effect. At around 1000\,K, monatomic potassium gas and KCl are supposed to be equally important potassium bearers, with decreasing temperature favoring KCl more \citep{lodders1999alkali,lodders&fegley2006chemistry_substellar}.


%
\subsection{15300--15800\,\AA{}  feature}

We identify a sharp feature at 15720\,\AA{} within a broader absorption feature at 15300--15800\,\AA{} (Fig.\ \ref{HCN_H2S}) that we cannot assign exactly to any molecule in the opacity database of the Data \& Analysis Center for Exoplanets (DACE)\footnote{\url{https://dace.unige.ch}}. However, we report the fact that HCN and H$_2$S have high opacity at these wavelengths, according to the DACE database, and laboratory transmission absorption spectra obtained using gas chambers (Fig.~\ref{HCN_H2S}) similar to the work by \citet{valdivielso2010gascell,martin2021ch4nh3}.
H$_2$S has been already detected in a T dwarf atmosphere via its sharp feature at 15900\,\AA{} \citep{tannock2022h2s_h2_T6}. Molecular hydrogen H$_2$ is the most abundant species but is unlikely to contribute to this feature via its weak molecular transition lines \citep{Roueff2019H2} or via its very broad collision-induced absorption \citep[CIA; ][]{richard2012H2CIA} in the dense atmosphere. Although we ruled out the possibility of telluric contamination by double checking the two $H$-band spectra of both the target and the telluric standard at the beginning and the end of the observations, we cannot discard that this feature is slightly blended with telluric bands or Brackett lines, which are densely populated there, hence weaker than seen.


\section{Radial Velocity and Galactic kinematics}
\label{esdT_WISE1810:kinematics}

The cross-correlation between the WISE1810 spectrum and that of a solar-metallicity RV standard is difficult because low metallicity fades out strong features (Fig.~\ref{fig:Klines}). 
We chose to use the \textit{crosscorrRV} function of PyAstronomy to individually cross-correlate three bands with the best-fit ATMO2020++ model. The wavelength of the model is in vacuum. We used a step $\Delta\mathrm{RV}$ of 20, 22, and 25 km\,s$^{-1}$ in the $JHK$ band, respectively, which is about a third of the resolution element size. Before the cross correlation, a 7th order polynomial was fitted and removed both from the spectrum and the model template. The 150 pixels of both ends of the spectra are softened by a cosine function to reduce the edge effect. We used a Gaussian function to fit the cross-correlation function (CCF), giving a window centred at the CCF peak with a width of $10 \times\Delta \mathrm{RV}$ km\,s$^{-1}$   (Fig.~\ref{CCF_ATMO}). We ran a Monte Carlo simulation to estimate the uncertainty of the radial velocity by adding random noise to each pixel and repeated for 10000 times. The noise of each pixel follows a Gaussian distribution with standard deviation inversely proportional to the S/N of the band.

We obtained mean barycentric corrections of each band using the mean time of the observations and the \textit{helcorr} function \citep{piskunov2002helcorr} in PyAstronomy, with the Earth rotation taking into account (Table~\ref{CCF_result}). The barycentric correction difference during the whole five-night observations is less than 2 km\,s$^{-1}$, which is included in the cross correlation process using the coadded spectra. We weighted the three measurements in Table~\ref{CCF_result} by the inverse of their variance and get the mean barycentric corrected heliocentric velocity $v_\mathrm{h}=-82.8\pm13.3$ km\,s$^{-1}$. This value is consistent at 3-$\sigma$ with the value of $-45.6\pm3.5$ km\,s$^{-1}$ obtained by \citet{lodieu2022W1810} but differs with a significance of 2-$\sigma$. The latter used the Cs{\small{I}} line at 8943\,\AA{} in the low-resolution GTC/OSIRIS spectrum and water lines in the $YJ$ band in the low-resolution GTC/EMIR spectrum. The fidelity of this measurement could be strongly hampered by the low S/R of the line, the telluric water vapor, and insufficient resolution. 


Based on the formulation of \citet{johnson1987uvw}, we used \textit{gal\_uvw} task in PyAstronomy to calculate the Galactic space velocity component $(U,V,W)$ of WISE1810 with the new heliocentric velocity value. Positive $U$, $V$,and $W$ means toward the Galactic center, in the direction of Galactic rotation, and toward the North Galactic Pole, respectively. Since WISE1810 is located not far from both the Galactic and ecliptic planes, the adoption of new heliocentric velocity will only change the $U$ component significantly. Using the same proper motion and robust parallax measurement reported by \citet{lodieu2022W1810}, we get heliocentric $(U,V,W)_\mathrm{h}= (-70.6\pm12.5, -56.1\pm4.3,  +26.6\pm1.3)$ km\,s$^{-1}$. By correcting the velocity of the local standard of rest \citep[LSR, $(U,V,W)_\odot= (+8.50\pm0.29, +13.38\pm0.29,  +6.49\pm0.26)$\,km\,s$^{-1}$; ][]{coskunoglu2011lsr}, we have $(U,V,W)= (-62.1\pm12.5, -42.7\pm4.4,  +33.1\pm1.4)$\,km\,s$^{-1}$ for WISE1810. These values suggest that WISE1810 is more likely associated with the thick disk rather than the halo, despite its very low metallicity. However, the possibility of thin disk membership cannot be entirely ruled out, see Fig.\,12 of \citet{lodieu2022W1810} and Fig.\,17 of \citet{zhang2017six_sdL_classification}.

%
%
\section{Conclusions}
\label{esdT_WISE1810:conclusions}
%


The improved NIR spectrum provides better insights into WISE1810, the nearest metal-poor T dwarf. 
CH$_4$ is clearly detected in the $HK$ bands while CO is below our detection limit, which implies a cool temperature ($\lesssim1200$\,K) and supports the T-type classification of WISE1810.  The best-fit ATMO2020++ model provides an effective temperature of $T_{\mathrm{eff}}\,=\,1000\pm100$\,K, a surface gravity of $\log g=5.5\pm0.5$\,dex, and a carbon abundance of [C/H]\,$=-1.5\pm0.2$\,dex constrained by the $17\pm6$\,ppm of atmospheric CH$_4$.  
We conclude that WISE1810 would have [Fe/H]\,$=-1.7\pm0.2$\,dex by adopting [C/Fe]\,$\sim0.2$\,dex \citep{nissen2014C_O_population}. Alkali lines (K{\small{I}}) are at least 25 to 60 times weaker than those of solar-metallicity early-T dwarfs. We provide an updated RV measurement of $-83\pm13$\,km\,s$^{-1}$, yielding a Galactic velocity $(U,V,W)= (-62.1\pm12.5,   -42.7\pm4.4,  +33.1\pm1.4)$\,km\,s$^{-1}$ with respect to the LSR, hence making WISE1810 very likely a thick disk member.

\section*{acknowledgments}
The authors thank the referee who provided insightful comments for this research.
The authors thank GTC telescope operators and support astronomers for their assistance in the acquisition of the data reported here.
JYZ, NL, VJSB and MRZO acknowledge support from the Agencia Estatal de Investigaci\'on del Ministerio de Ciencia, Innovaci\'on y Universidades under grants PID2019-109522GB-C53, PID2022-137241NB-C41 (JYZ, NL, VJSB), and PID2022-137241NB-C42 (MRZO)\@.
JYZ, EGE and NV were funded for this research by the European Union ERC AdG SUBSTELLAR grant agreement number 101054354\@.
BG acknowledges support from the Polish National Science Center (NCN) under SONATA grant No. 2021/43/D/ST9/0194\@.
Based on observations made with the Gran Telescopio Canarias (GTC), in the Spanish Observatorio del Roque de los Muchachos of the Instituto de Astrof\'isica de Canarias, on the island of La Palma, under program GTC73-24A (PI Lodieu).
EMIR has been funded by GRANTECAN S.L.\ via a procurement contract; by the Spanish funding agency grants AYA2001-1656, AYA2002-10256-E, FIT-020100-2003-587, AYA2003-01186, AYA2006-15698-C02-01, AYA2009-06972, AYA2012-33211, AYA2015-63650-P and AYA2015-70498-C2-1-R; and by the Canarian funding agency grant ACIISI-PI 2008/226.
This research has made use of the Simbad and Vizier databases, operated at the centre de Donn\'ees Astronomiques de Strasbourg (CDS), and of NASA's Astrophysics Data System Bibliographic Services (ADS).
This publication makes use of data products from the Two Micron All Sky Survey, which is a joint project of the University of Massachusetts and the Infrared Processing and Analysis Center/California Institute of Technology, funded by the National Aeronautics and Space Administration and the National Science Foundation.
This research has made use of the Spanish Virtual Observatory (https://svo.cab.inta-csic.es) project funded by MCIN/AEI/10.13039/501100011033/ through grant PID2020-112949GB-I00.

%

\vspace{5mm}
\facilities{GTC}

\software{IRAF \citep{tody1986iraf,tody1993iraf}, PyEMIR\footnote{\url{https://pyemir.readthedocs.io/}}, Astrometry.net \citep{dustin2010astrometry.net},  Photutils \citep{photutil}, Astropy \citep{astropy2013,astropy2018,astropy2022}, PyAstronomy\footnote{\url{https://github.com/sczesla/PyAstronomy}} \citep{czesla2019pyastronomy}}



\clearpage
\appendix
\section{Observation Logs}
%
%
\begin{table}[htbp]
\centering
\footnotesize
 \caption{Summary of GTC/EMIR observations.}
 \begin{tabular}{cccccc}
 \hline
 \hline
Target &  
MJD &  Exp.  & Seeing & Fil/Grism & Std.  \cr
 \hline
WISE1810  
&60512.90 & 7$\times$10s  & 0\farcs6 & $J$  &  \cr
WISE1810  
&60512.91 & 7$\times$10s  & 0\farcs6 & $K_{s}$  &  \cr
WISE1810  
&60512.91 & 7$\times$10s  & 0\farcs6 & $H$  &  \cr
%
WISE1810  
&60513.03 & 24$\times$360s  & 0\farcs6 & spec $J$ & HIP 88374  \cr
WISE1810  
&60514.00 & 36$\times$360s  & 0\farcs5 & spec $H$ & HIP 88374 \cr
WISE1810  
&60515.00 & 36$\times$360s  & 0\farcs5 & spec $K$ & HIP 88374 \cr
WISE1810  
&60516.00 & 36$\times$360s  & 0\farcs5 & spec $K$ & HIP 88374 \cr
WISE1810  
&60517.00 & 12$\times$360s  & 0\farcs5 & spec $J$ & HIP 88374 \cr
2MASS2030  
&60517.10 & 4$\times$200s  & 0\farcs5 & spec $J$ & HIP 88734 \cr
\hline
 \label{tab_WISE1810:tab_logs}
 \end{tabular}
\end{table}

\section{Spectrum}
\begin{figure}[htbp]
    \centering
    \includegraphics[width=0.85\linewidth]{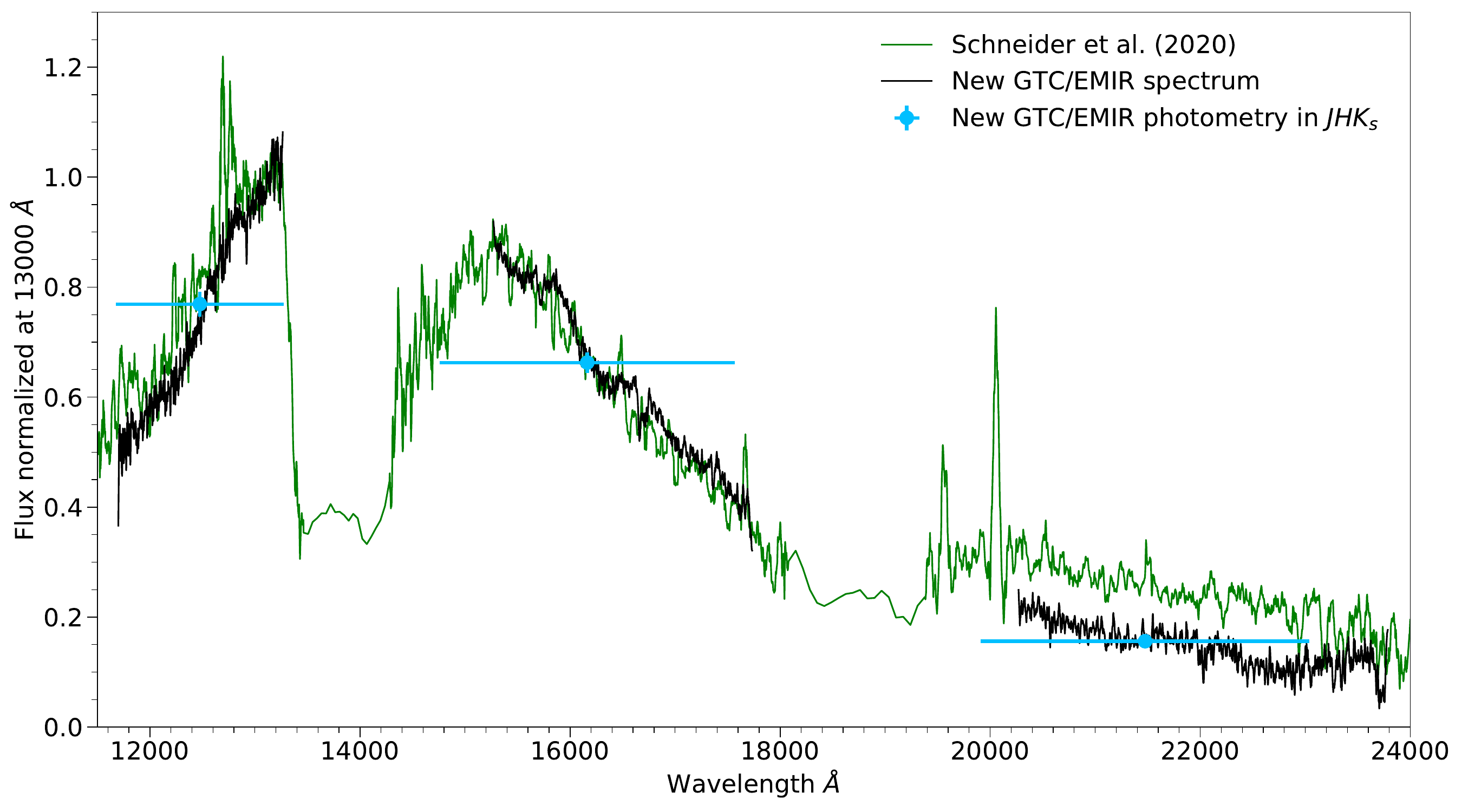}
    \caption{Comparison between old spectrum from \citet{Schneider2020W0414_W1810} and new GTC/EMIR spectrum. While GTC/EMIR photometry agrees with photometry from \citet{Schneider2020W0414_W1810}, the two spectra are not matched.}
    \label{compare}
\end{figure}

\clearpage
\begin{figure}[htbp]
    \centering
    \includegraphics[width=0.45\linewidth]{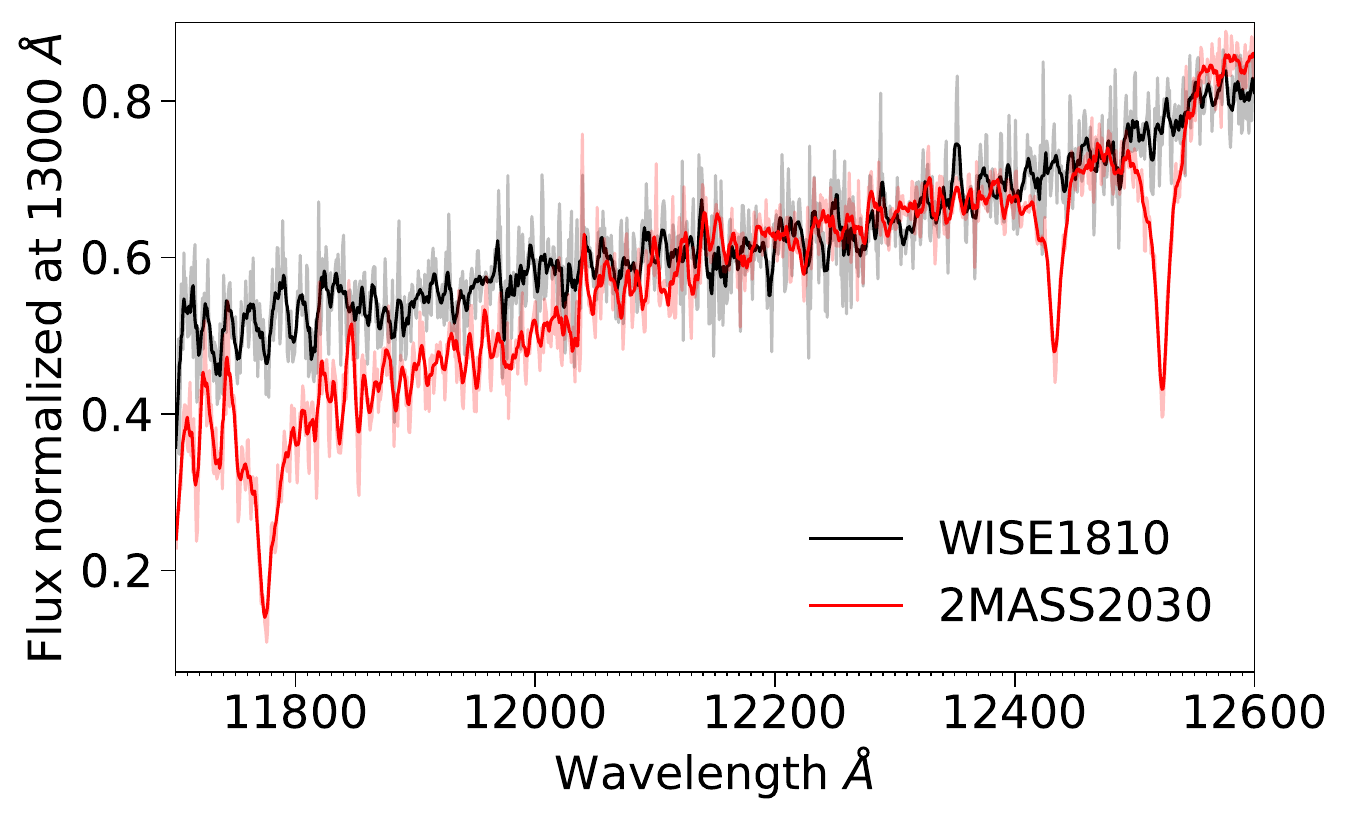}
    \caption{Comparison between WISE1810 and the 2MASS2030 T1.5 dwarf in the region of three K{\small{I}} lines. The lighter coloured curves are original data and the dark ones are smoothed by a factor of 7\@.} 
    \label{fig:Klines}
\end{figure}

 \begin{figure}[htbp]
    \centering
    \includegraphics[width=0.5\linewidth]{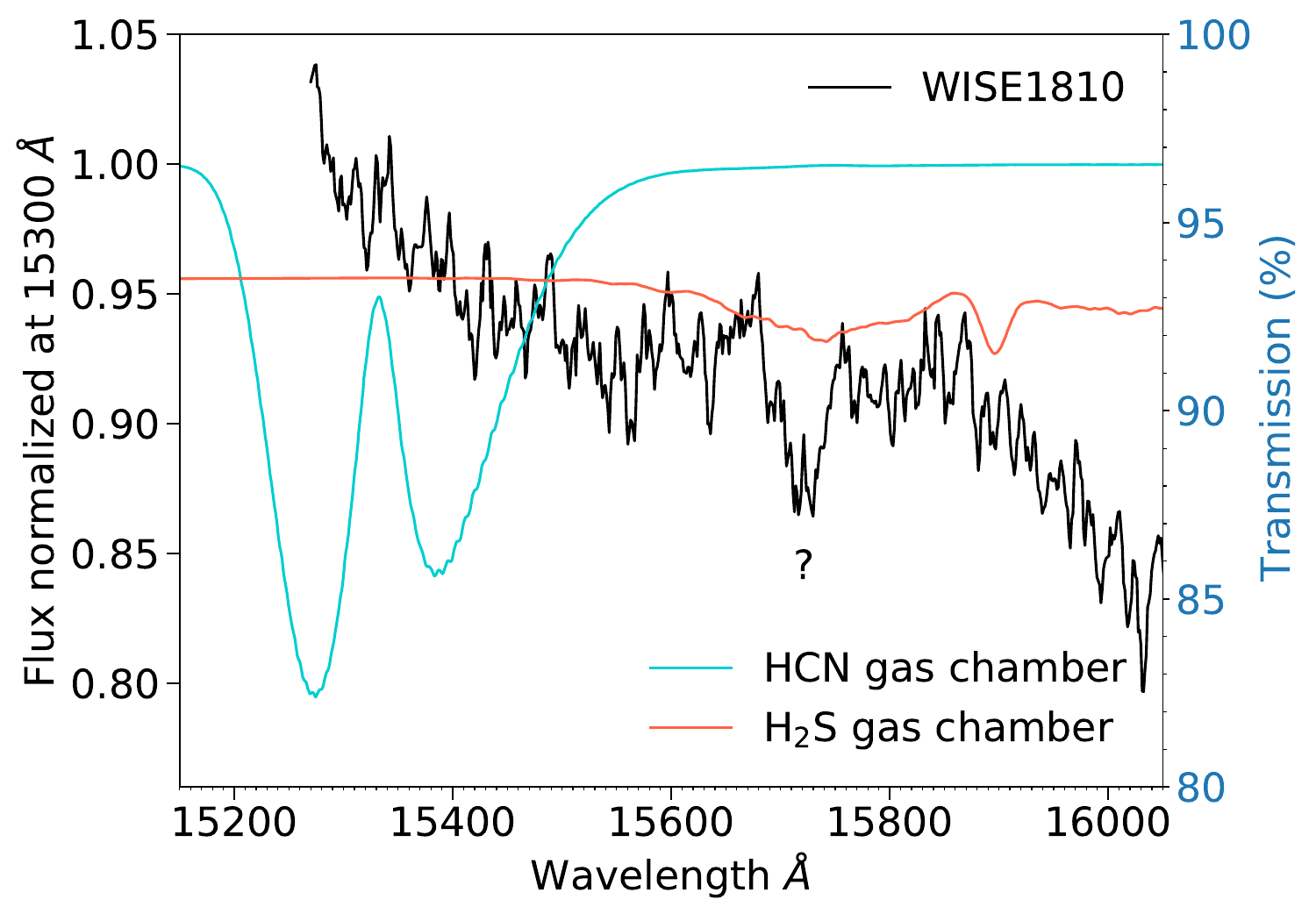}
    \caption{15300-15800\,\AA{} feature compared with transmission absorption spectra of HCN and H$_2$S obtained using gas chamber under room temperature and 1 atm pressure.}
    \label{HCN_H2S}
\end{figure}

\section{Chemical abundance profiles}
\begin{figure}[htbp]
    \centering
    \includegraphics[width=0.45\linewidth]{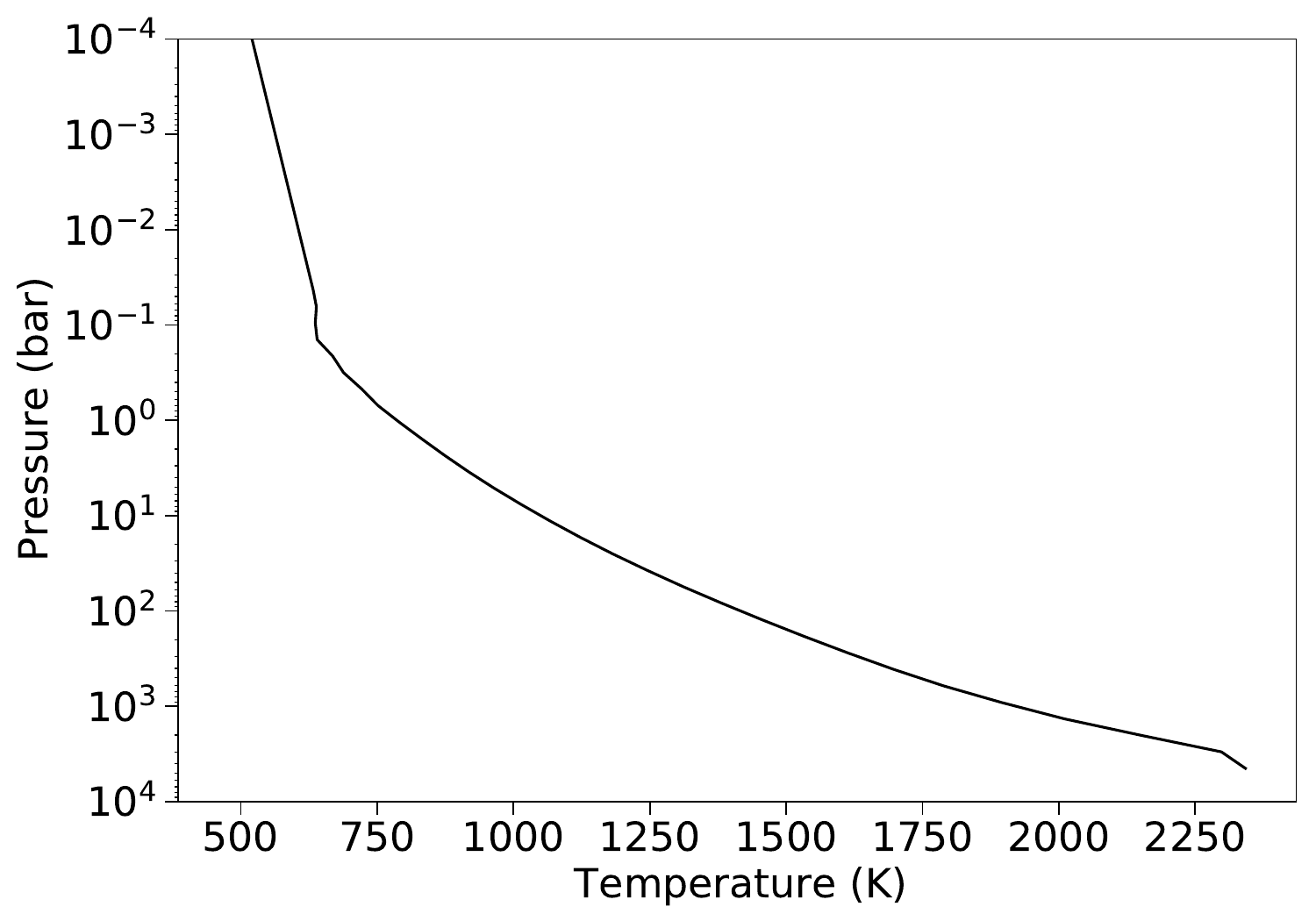}
    \includegraphics[width=0.45\linewidth]{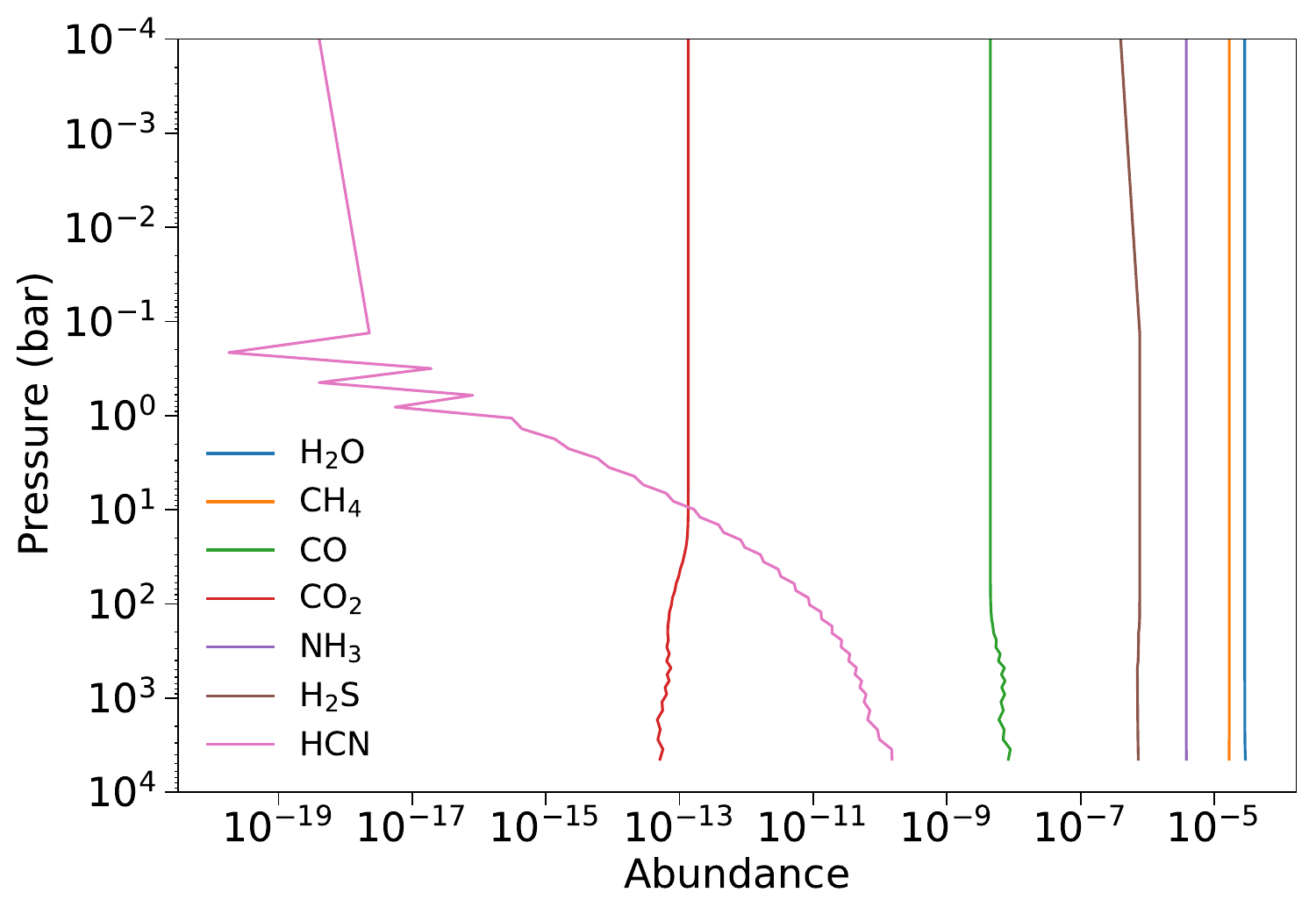}
    \caption{Temperature-pressure profile (left) and abundance-pressure profiles of most frequent molecules (right) in the atmosphere of WISE1810 from the adjusted ATMO2020++ model with non-equilibrium chemistry.}
    \label{pressure-abundance}
\end{figure}

\clearpage
\section{Radial velocity cross correlation}
\begin{figure}[htbp]
    \centering
    \includegraphics[width=0.325\linewidth]{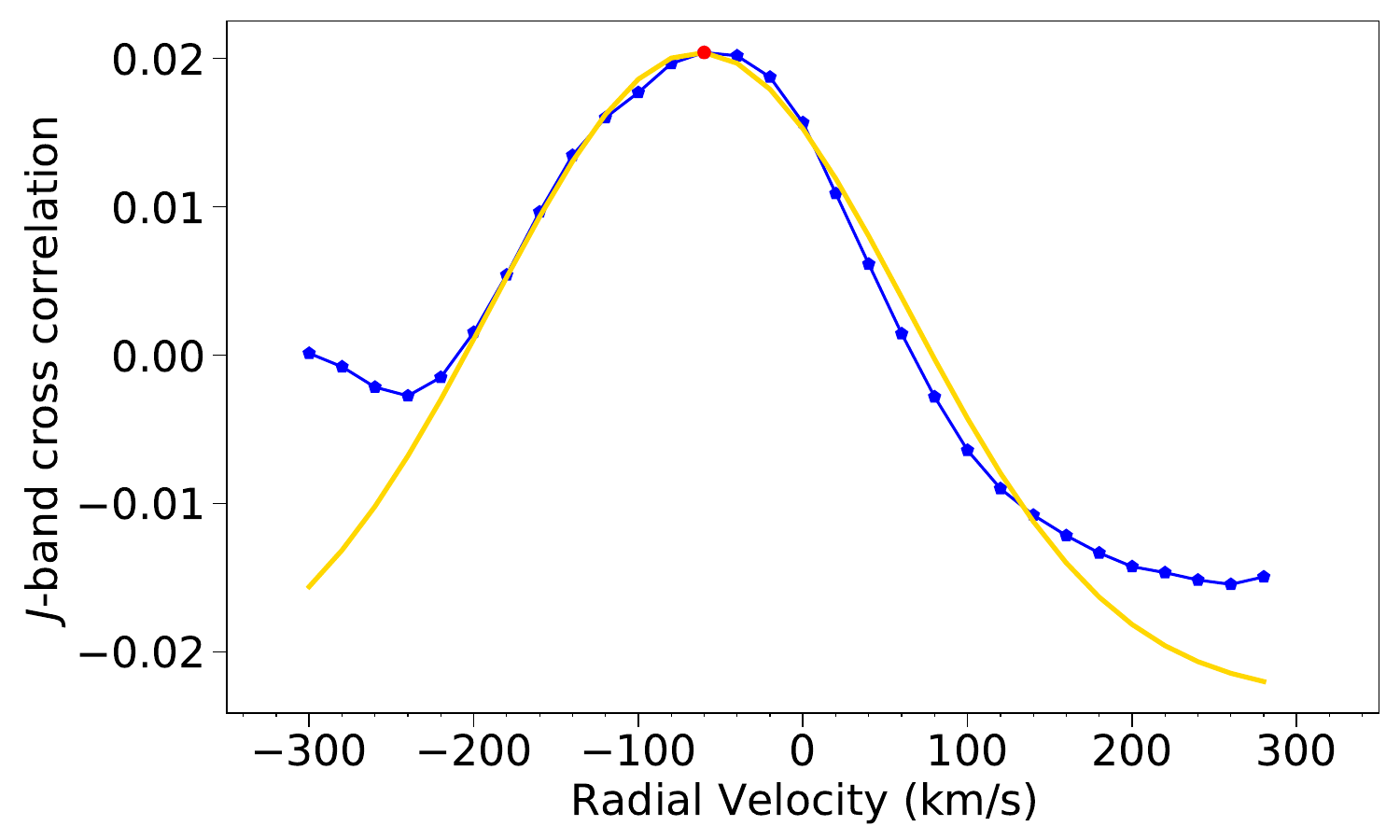}
    \includegraphics[width=0.325\linewidth]{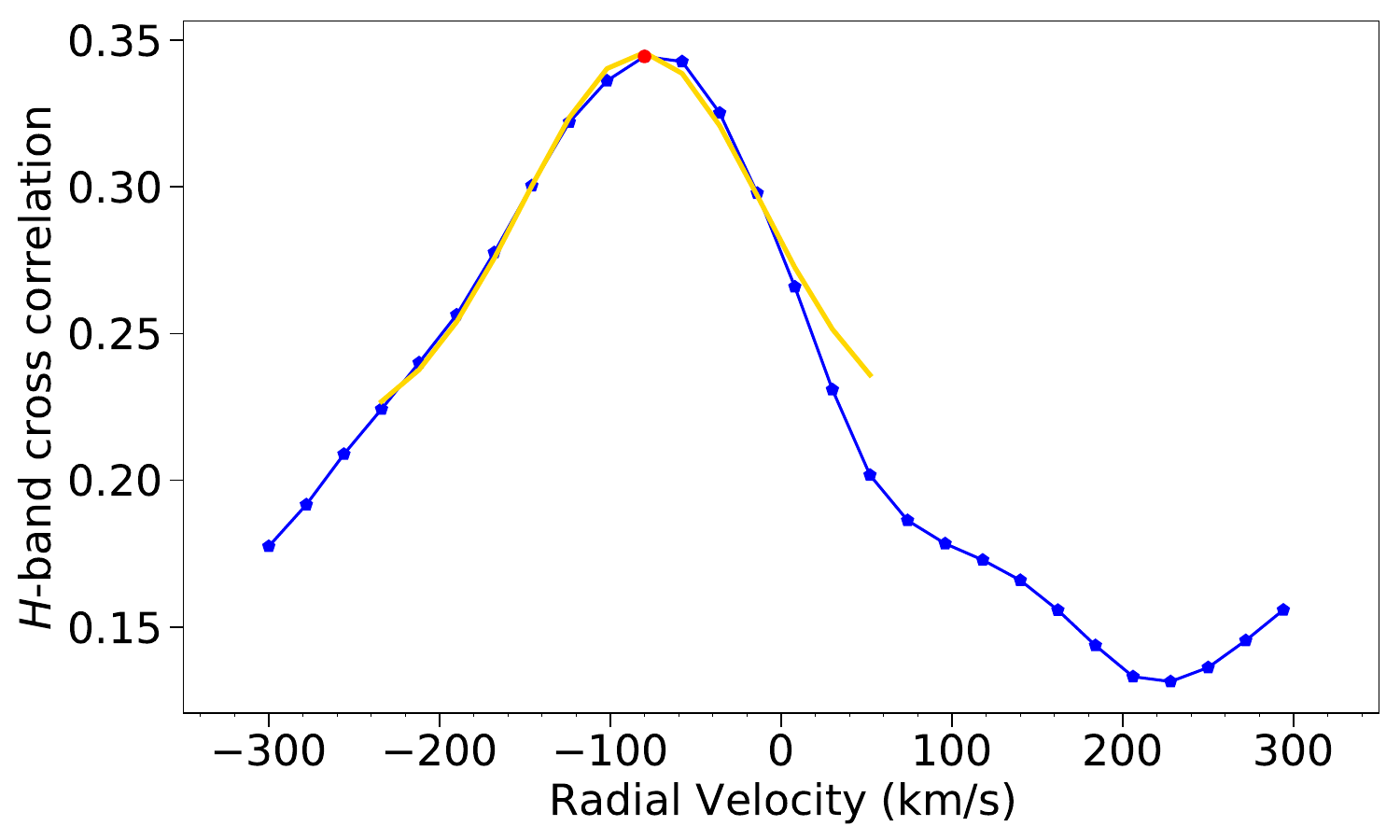}
    \includegraphics[width=0.325\linewidth]{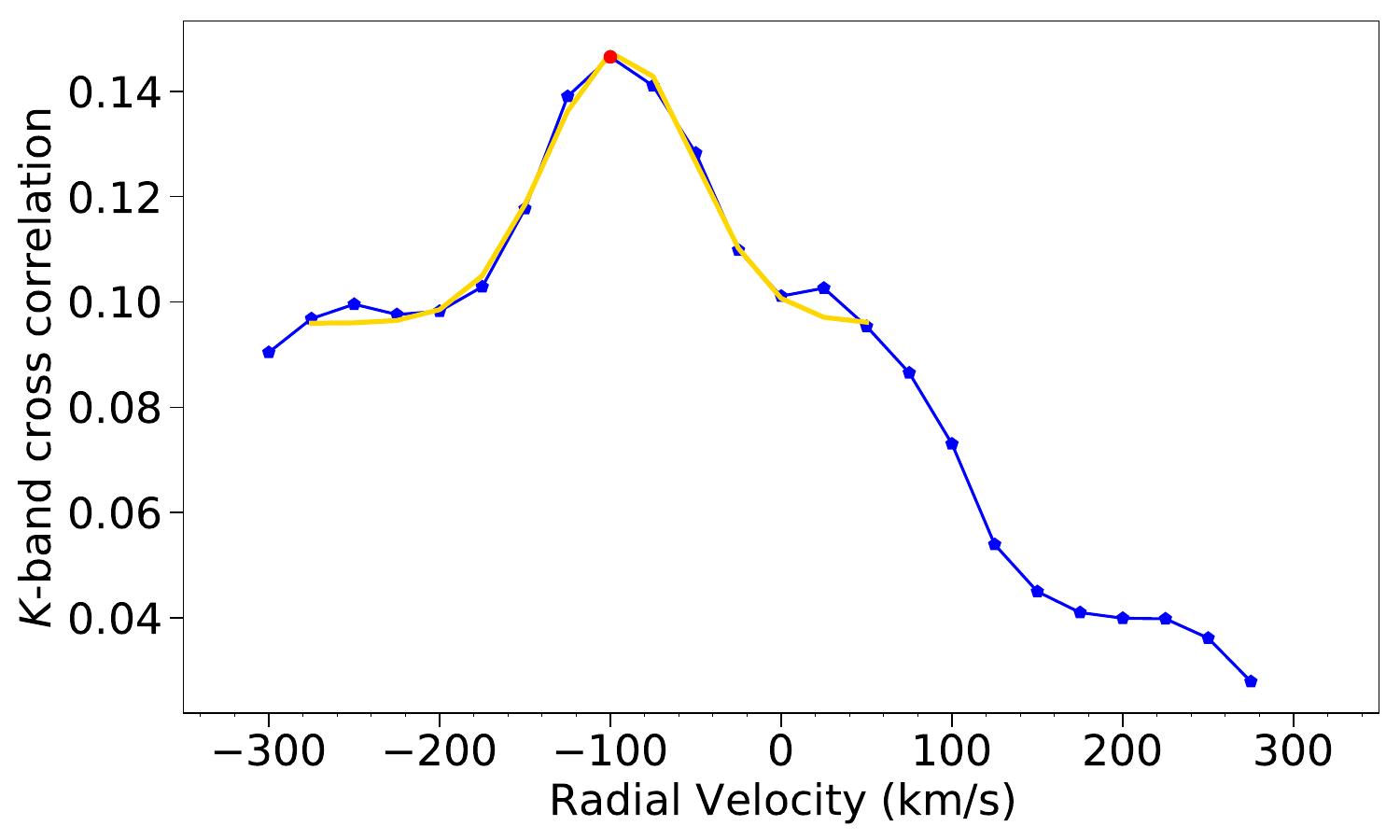}
    \caption{Normalized CCFs against radial velocity (in blue) with their maximum (red dot) and Gaussian fits (in gold) in the $JHK$ bands.}
    \label{CCF_ATMO}
\end{figure}

\begin{table}[htbp]
    \centering
    \footnotesize
    \caption{RV of WISE1810 from each band.}
    \begin{tabular}{ccccc}
     \hline
 \hline
         Band & RV (km\,s$^{-1}$) & Bary. corr. (km\,s$^{-1}$)  & $v_\mathrm{h}$ (km\,s$^{-1}$)  \\
         \hline 
         $J$ & $-71.4\pm15.2$ & $13.9$ & $-85.3\pm15.2$ \\
         $H$ & $-66.7\pm9.2$ & $13.5$ & $-80.2\pm9.2$ \\
         $K$ & $-93.3\pm35.8$ & $13.9$ & $-107.2\pm35.8$ \\
         \hline
    \end{tabular}
    \label{CCF_result}
\end{table}


\bibliography{bibliography}{}
\bibliographystyle{aasjournal}


\end{CJK*}
\end{document}